\documentclass[11pt,letterpaper]{article}

\usepackage[utf8]{inputenc}
\usepackage{amsmath,amssymb,pdfpages,geometry,nicematrix,graphicx,placeins, mathtools, tikz-cd, upgreek, booktabs, multirow} 
\usepackage[hidelinks]{hyperref}
\usepackage{appendix}
\usepackage{float}

\usepackage{microtype}

\usepackage[labelfont=bf]{caption}
\usepackage[style=nature]{biblatex}
\AtEveryBibitem{
    \clearfield{urlyear}
}

\usepackage{pgfplotstable}
\pgfplotsset{compat=1.17} 

\renewcommand{\vec}[1]{\mathrm{\mathbf{#1}}}

\title{De novo structural ensemble determination from single-molecule X-ray scattering: A Bayesian approach}
\author{Steffen Schultze$^1$, Helmut Grubmüller$^{2*}$}
\date{%
    \textit{Max Planck Institute for Multidisciplinary Sciences}\\[2ex]
    $^1$sschult@mpinat.mpg.de \,\,\,$^2$hgrubmu@gwdg.de\\[2ex]
    \today
}

\begin{document}

\maketitle

\begin{abstract}
Single molecule X-ray scattering experiments with free electron lasers have opened a new route to the structure determination of biomolecules. 
Because typically only very few photons per scattering image are recorded and thus the signal-to-noise ratio is very low in this extreme Poisson regime, structure refinement is quite challenging. 
In addition, in each scattering event the orientation of the biomolecule is random and unknown. 
As a further layer of complexity, many biomolecules show structural heterogeneity and conformational transitions between different distinct structures; these structural dynamics are averaged out by existing refinement methods. 
To overcome these limitations, here we developed and tested a rigorous Bayesian approach and demonstrate that it should be possible to determine not only a single structure, but an entire structural ensemble from these experiments. 
Using $10^6$ synthetic scattering images generated from molecular dynamics trajectories, our approach was able to resolve ensembles of eight alanine dipeptide conformers at $2\,${\AA} resolution; similarly, we determined the unfolded ensemble of the protein chignolin at $4-7\,${\AA} resolution using $1.2\cdot 10^7$ images.
Unexpectedly, much fewer images are required to determine an ensemble of $n$ structures of $m$ atoms each than a single structure of $n\times m$ atoms, 
i.e., of the same total number of degrees of freedom. These findings show that X-ray scattering experiments using state-of-the-art free electron lasers should allow one to determine not only biomolecular structures, but whole structure ensembles and, ultimately, `molecular movies'.
\end{abstract}

\newpage
\section{Introduction}
Ultrashort pulse X-ray scattering experiments offer the possibility to take 'snapshots' of biomolecular structures with atomistic 
spatial and femtoseconds time resolution~\cite{hajdu_single-molecule_2000, huldt_diffraction_2003, gaffney_imaging_2007, miao_beyond_2015}. 
Still, most current experiments focus on nano-crystals~\cite{chapman_femtosecond_2006, chapman_femtosecond_2011, boutet_high-resolution_2012, fromme_femtosecond_2011, kirian_femtosecond_2010, schlichting_serial_2015, roedig_high-speed_2017, barends_serial_2022}. 
Like classical X-ray crystallography, these average over many molecules and, therefore, time resolved structure determination requires strict synchronization, typically by optical laser pulses~\cite{barends_serial_2022}. 
Scattering on single particles or molecules avoids this limitation and should enable us to
advance towards structure ensembles and, ultimately, time resolved conformational and functional motions without the need for synchronization \cite{ourmazd_cryo-em_2019,van_thor_advances_2019}.

\begin{figure}
    \centering
    \includegraphics[height = 0.3\textwidth]{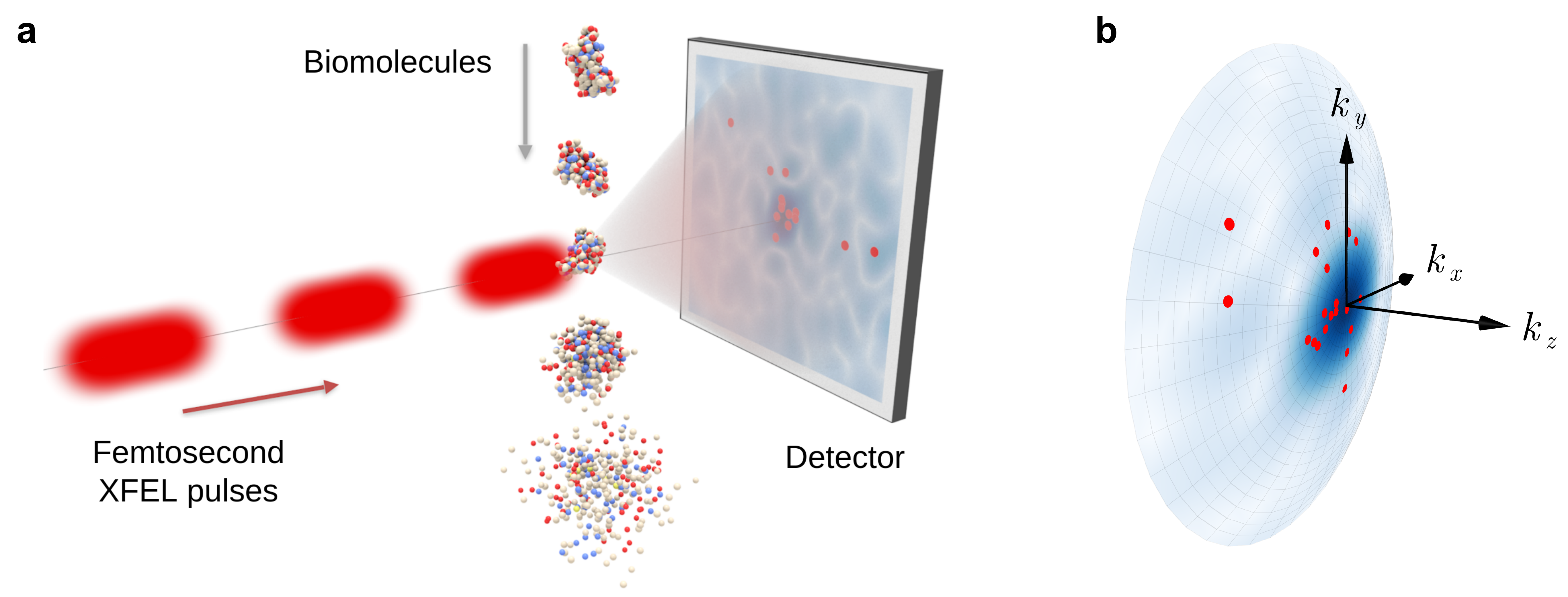}
    \caption{Single molecule scattering experiment. \textbf{a} A stream of single molecules is hit by femtosecond X-ray pulses, and the scattered photons are recorded as images (image reproduced from von Ardenne et al.~\cite{von_ardenne_structure_2018}). \textbf{b} The scattered photons (red dots) are distributed on the Ewald sphere according to the 3D-intensity function $I$ (blue).}
    \label{fig: experiment}
\end{figure}

In such `hit and destroy' experiments, a stream of single molecules is exposed to a beam of high intensity femtosecond X-ray free electron laser (XFEL) pulses (Fig.~\ref{fig: experiment}a). For each hit the positions of the scattered photons (red dots) on the detector are recorded as a scattering image~\cite{neutze_potential_2000}. 
Importantly, the ultra-short pulses serve to outrun the subsequent destruction of the particles due to radiation damage, but also
imply that only very few photons are being recorded for each molecule~\cite{gaffney_imaging_2007}.

The feasibility of this approach has already been demonstrated by a number of experiments~\cite{seibert_single_2011, ekeberg_three-dimensional_2015, hosseinizadeh_high-resolution_2014}, but so far only structures of relatively large specimen at low resolution have been successfully determined, for instance of entire mimivirus particles~\cite{seibert_single_2011, ekeberg_three-dimensional_2015}  ($450\,\text{nm}$ in diameter) and coliphage viruses~\cite{hosseinizadeh_high-resolution_2014} ($20\,\text{nm}$ in diameter). 
Whereas for large specimens many photons are scattered per image, for example $10^7$ for the mimivirus~\cite{seibert_single_2011, ekeberg_three-dimensional_2015}, for typical proteins only 10-100 coherently scattered photons per image are expected~\cite{yoon_comprehensive_2016, hantke_condor_2016}, which further complicates structure determination particularly for small molecules. 
Such images can be obtained with an intensity of $10^{12}$ photons per pulse at $5\,\mathrm{keV}$ and a $1\,\mathrm{\text{\textmu} m}$ beam diameter~\cite{von_ardenne_structure_2018}, for example from the XFELs at DESY or SLAC.

Most importantly, the orientation of the molecules at the time of scattering is typically unknown, which poses an additional and substantial refinement challenge.
These issues are particularly challenging for the structure refinement of small specimen such as proteins or protein complexes at near-atomic resolution. 
A number of methods have been proposed to address these issues, such as orientation determination methods~\cite{shneerson_crystallography_2008,loh_reconstruction_2009, walczak_bayesian_2014,kassemeyer_optimal_2013,elser_three-dimensional_2011, tegze_atomic_2012, donatelli_reconstruction_2017, flamant_expansion-maximization-compression_2016, ayyer_dragonfly_2016} and manifold embedding algorithms~\cite{fung_structure_2009,moths_bayesian_2011,schwander_symmetries_2012,giannakis_symmetries_2012}, which, however, typically require 100 to 1000 photons per scattering image. 
As an alternative, correlation based approaches~\cite{saldin_structure_2009, saldin_beyond_2010, saldin_new_2011, saldin_structure_2010,kurta_correlations_2017} have recently allowed substantial advancements and have been shown 
to require, quite counterintuitively, only three photons per image~\cite{von_ardenne_structure_2018}. However, as for all other approaches proposed so far, it is impossible to systematically include shot noise, incoherently scattered photons, background scattering, or detector noise in the structure refinement.

Finally, many biomolecules show structural heterogeneity and conformational dynamics between different distinct structures, which, when resolved, would provide a direct view on biomolecular function. 
Hence, and similar to the current main challenge in cryogenic electron microscopy~\cite{guaita_recent_2022}, not one but many structures need to be extracted from the scattering images. 
In this scenario, in addition to the orientation, also the current conformer for each scattering image is unknown. 
Whereas both orientation determination methods and manifold-based methods have been applied to determine multiple conformational structures~\cite{zhuang_unsupervised_2022,hosseinizadeh_conformational_2017-1}, the required large number of photons per image precludes their application to single biomolecules.

To overcome these issues, we developed and assessed a rigorous Bayesian method for multiple structure determination from single-molecule scattering images. 
We will demonstrate that this method can not only determine a single structure, but also ensembles consisting of multiple structures at high resolution.
Unexpectedly, much fewer images are required for refining an ensemble of $n$ conformers consisting of $m$ atoms each than for refining a single structure consisting of $m\times n$ atoms, which should render biomolecular ensemble determination accessible to state-of-the-art experiments.

\section{Results}

\paragraph{Summary of the approach.}

For each scattering image, the positions of the recorded photons specify vectors $\vec k_1, \dots, \vec k_l$ on the Ewald sphere in Fourier space (red dots in Fig.~\ref{fig: experiment}b). The probability of observing a photon at a particular position on the detector is proportional to the 3D intensity function $I(\vec k) \propto |\mathcal F\{\vec R\rho_i\}(\vec k)|^2$ at the corresponding position $\vec k$ on the Ewald sphere, which in turn is given by the Fourier transform of the electron density $\rho_i$ of conformer $i$. Here, $\vec R$ is the unknown orientation of the molecule for this particular image.

It follows that the probability of observing an image with photon positions $\vec k_1, \dots, \vec k_l$ is obtained by averaging over both the conformational ensemble $\boldsymbol\uprho = \{\rho_1, \dots, \rho_n\}$ with weights $\vec w = \{w_1, \dots, w_n\}$ as well as over all orientations $\vec R$. Because the scattering images are statistically independent from each other, the total probability of observing the complete set of all images $\mathcal I$ reads
\begin{equation}
    P(\mathcal I \,|\, \boldsymbol\uprho, \vec w) \propto \prod_{(\vec k_1, \dots, \vec k_l) \in \mathcal I} \sum_{i=1}^n w_i\int_{\mathrm{SO}(3)}P(\vec k_1, \dots, \vec k_l \,|\, \vec R\rho_i)\, \mathrm d \vec R\,.
    \label{eq: likelihood}
\end{equation}
This probability serves to determine either a single structure or a structural ensemble by sampling from the Bayesian posterior probability $P(\boldsymbol\uprho, \vec w \,|\, \mathcal I) \propto P(\mathcal I \,|\, \boldsymbol\uprho, \vec w)P(\boldsymbol\uprho, \vec w)$ using a Markov chain Monte Carlo approach. For the prior $P(\boldsymbol\uprho, \vec w)$ the orientations are assumed to be uniformly distributed. To minimize the number of required degrees of freedom, and as a means of regularization, we chose a physically motivated representation of each $\rho_i$ in terms of a sum of Gaussian functions, which also completes the definition of the prior.


For a typical protein consisting of 50 to several hundred residues, the number of required degrees of freedom remains large and poses a formidable sampling challenge. 
To address this issue, we have implemented a hierarchical simulated annealing approach. 
Starting at very low resolution, the macromolecular structures were sampled in multiple hierarchical stages of increasing resolution. To increase the sampling efficiency, in each of these stages, for each Markov step the previous ensemble of structures of maximal posterior probability was used as a proposal density.
To this end, the scattering images that would have been observed for a smoothed low resolution copy of the original molecule were obtained from the original images by rejection sampling using the convolution theorem (see the Methods section). 

Further, we adapted the Bayesian formalism such that only those images are used which contain new information, that is, photons for which the magnitude of $\vec k$ is larger than the threshold of the resolution from the previous stage.
With increasing resolution, the fraction of such useful images becomes very small, thus enhancing computational efficiency up to two orders of magnitude. The approach is described in detail in the Supplementary Information.

\begin{figure}[t]
    \centering
    \includegraphics[width = \textwidth]{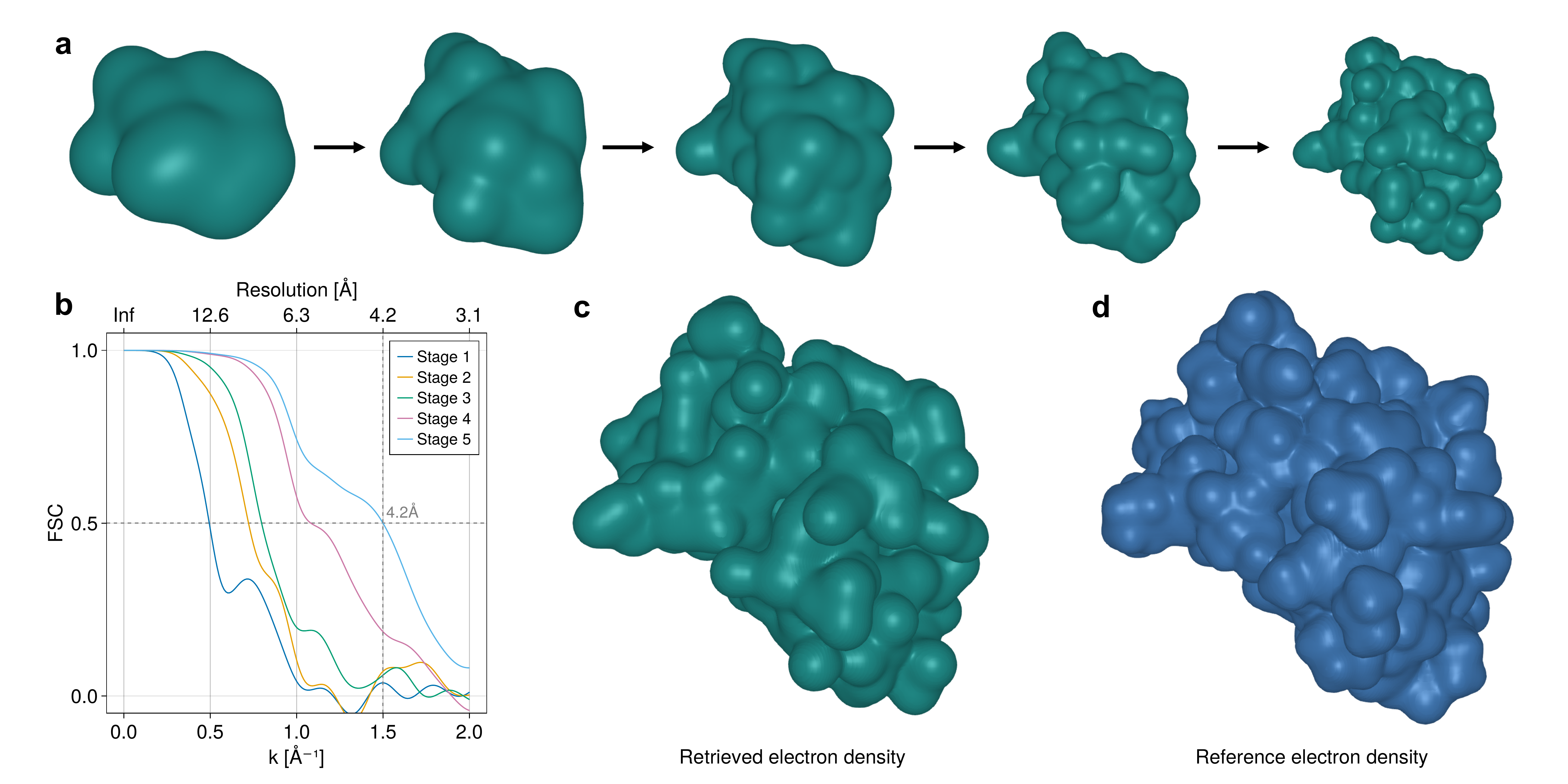}
    \caption{Structure determination of Crambin. \textbf{a} Hierarchical stages of retrieved electron densities. \textbf{b} Fourier shell correlation between the retrieved densities and the reference density. \textbf{c} Retrieved electron density. \textbf{d} Reference electron density.}
    \label{fig: crambin}
\end{figure}

\paragraph{Sample test refinements.}
Because our Bayesian approach uses all available information, we expect it to require fewer scattering images to achieve a certain resolution than, for example, correlation based methods.
To assess this aspect, we first tested our method on the single structure level, using the same 46-residue protein crambin~\cite{jelsch_accurate_2000} as in our previous study~\cite{von_ardenne_structure_2018}.
A total of $10^8$ noise-free synthetic images were generated, containing a realistic average of $15$ photons each. 
From these, the structure was solved in five hierarchical stages (Fig.~\ref{fig: crambin}a), increasing the number of degrees of freedom by a factor of two in each stage. 
For the final stage, a representation of $\rho$ consisting of $184$ Gaussian functions was used, which is four times the number of residues. For more details see Supplementary Note~1. 
Indeed, using only half of the total number of scattered photons, a similar Fourier shell correlation resolution~\cite{van_heel_fourier_2005} of $4.2\,${\AA} (Fig.~\ref{fig: crambin}b) is obtained as with the previous correlation based method~\cite{von_ardenne_structure_2018}.

\begin{figure}[t]
    \centering
    \includegraphics[width = \textwidth]{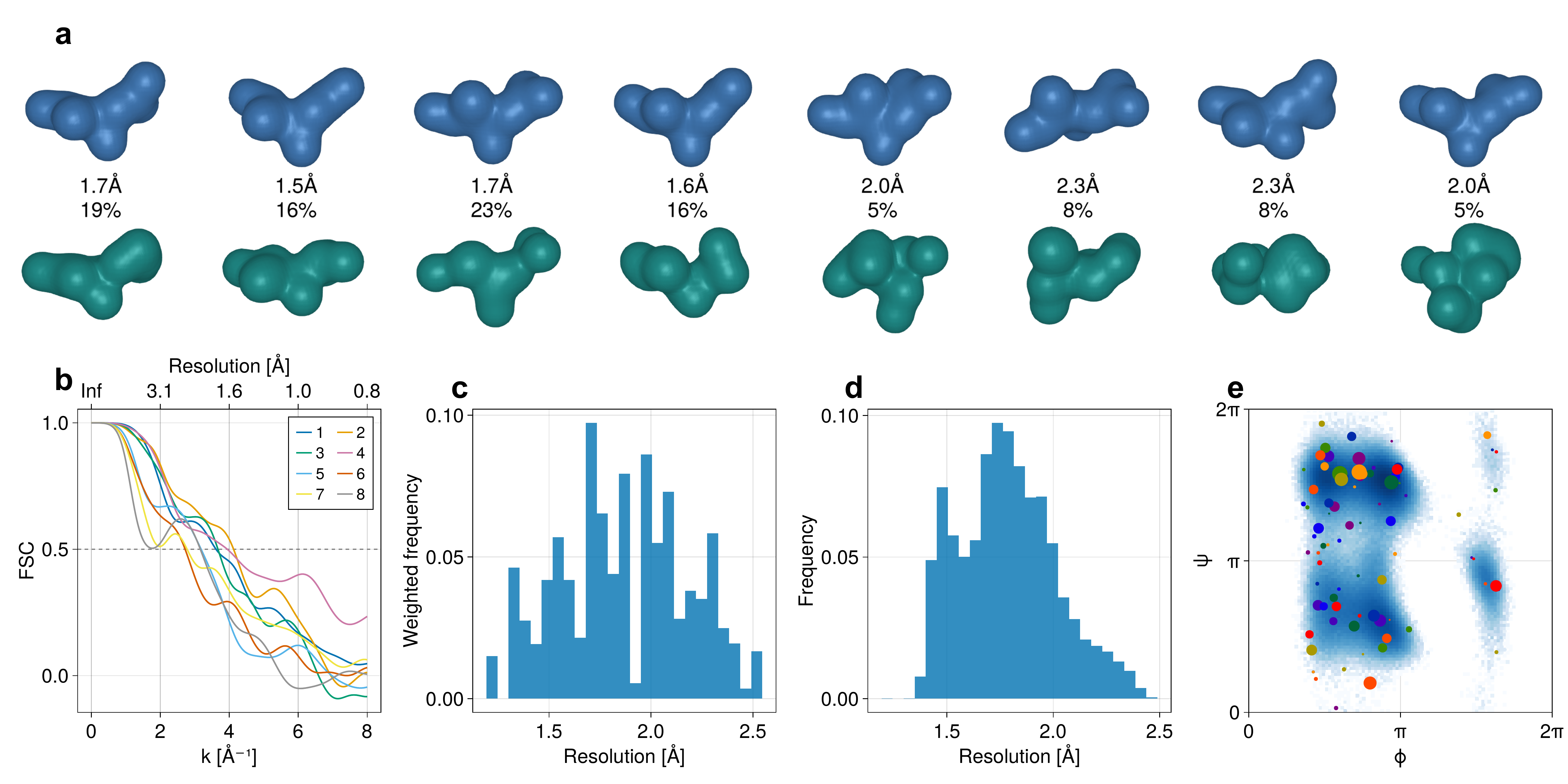}
    \caption{Structural ensemble determination of the alanine dipeptide. \textbf{a} Reconstructed conformers (green), the corresponding weights, and the nearest neighbors in the input trajectories (blue) with the corresponding resolutions. \textbf{b} Fourier shell correlations used to compute these resolutions (from left to right). \textbf{c} Weighted resolution distribution for $10$ independent runs from the the same data. \textbf{d} Resolution distribution over the time steps of the input trajectories relative to their nearest neighbors among the determined structures from all $10$ runs. \textbf{e} Ramachandran plot for the input trajectories (shown as a density) and the determined structures from all $10$ runs (points, the colors indicate the separate runs). }
    \label{fig: dipeptide}
\end{figure}

Next, to demonstrate that our method can resolve not only a single protein structure, but also ensembles of multiple conformers, we used three molecular dynamics trajectories of alanine dipeptide~\cite{scherer_pyemma_2015} of length $250\,\text{ns}$ each to generate $10^6$ scattering images, using a randomly chosen snapshot for each image. 
As before, an average of $15$ photons per image were generated. Using our approach,
a weighted ensemble of eight conformers was determined from these images (Fig.~\ref{fig: dipeptide}), with each conformer being described by a sum of $10$ Gaussian functions. 
To obtain sufficient statistics, a total of $10$ independent simulated annealing runs were carried out, using the same image set.

To assess the quality of the obtained structure ensemble, for each of the eight structures the resolution with respect to its nearest neighbor in the input trajectories was calculated using Fourier shell correlations~\cite{van_heel_fourier_2005} (Fig.~\ref{fig: dipeptide}b), resulting in a weighted average resolution of $1.8\,\text{{\AA}}$. 
This result shows that the obtained eight structures are indeed close to the reference ensemble. 
To also assess the accuracy of the entire ensemble, for each time step in the input trajectories, the resolution with respect to its nearest neighbor among all the determined structures was calculated (Fig.~\ref{fig: dipeptide}d). 
As a main result we found that $90\%$ of the input trajectories are within $2.1\,${\AA} Fourier shell correlation resolution of the determined structures, and that all of the trajectory frames are within $2.5\,${\AA} resolution of the determined structures, thus demonstrating atomistic resolution. 
Figure~\ref{fig: dipeptide}e compares the $10$ obtained ensembles with the reference ensemble using a Ramachandran plot~\cite{ramachandran_stereochemistry_1963} showing the distribution of the torsion angles $\phi$ and $\psi$. 
For each of the determined structures its nearest neighbor in the input trajectories was used to compute these angles. 
As can be seen, the reference density is well represented by the determined structures. 

\begin{figure}[t]
    \centering
    \includegraphics[width = \textwidth]{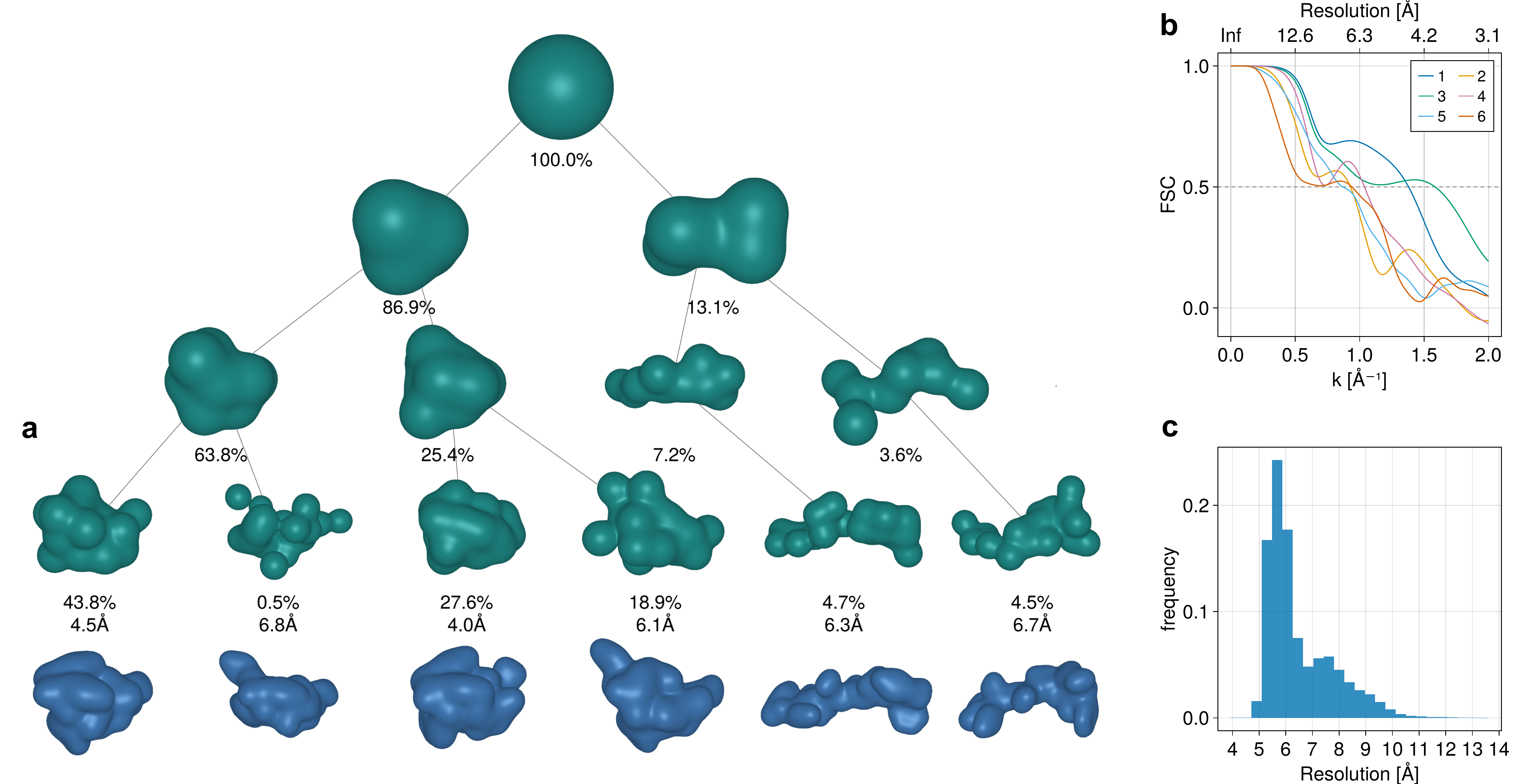}
    \caption{Structural ensemble determination for chignolin. \textbf{a} Hierarchical stages of retrieved structures (green) and their nearest neighbors (blue) in the input trajectories with the corresponding resolutions. \textbf{b} Fourier shell correlations of the reconstructed structures relative to their nearest neighbors (from left to right). \textbf{c} Resolution distribution over the time steps of the input trajectories relative to their nearest neighbors among the determined structures. }
    \label{fig: 5awl}
\end{figure}

Next, we asked if our method is also capable of extracting structural ensembles for the larger mini-protein chignolin~\cite{honda_crystal_2008}, comprising $10$ residues. 
To that end, $50$ molecular dynamics trajectories of length $10\,\text{\textmu s}$ were used to generate $1.2\cdot10^7$ images with, on average, $15$ photons each. 
As a further challenge, this ensemble also contained unfolded structures. 
From the obtained images, we determined multiple stages of weighted structural ensembles of increasing resolution and increasing number of conformers (Fig.~\ref{fig: 5awl}a). 
As above, resolutions were computed using Fourier shell correlations (Fig.~\ref{fig: 5awl}b,c), finding a weighted average resolution of $4.7\,${\AA} for the folded conformers, and $6.4\,${\AA} for the unfolded conformers. 
Interestingly, in the final stage one of the six determined weights is nearly zero, suggesting that five conformers suffice for the used number of images at this resolution level. 
It is also worth noting that the $9\%$ fraction of unfolded states in the reference structure ensemble was correctly identified.

\paragraph{Scaling.}
For both of the above sample applications described above we observed, unexpectedly, that resolving $n$ conformational structures consisting of $m$ residues each required much fewer scattering images and photons than resolving a single $n\times m$ residue structure of the same total size and complexity --- even in cases where the conformers of the ensemble are very different from each other.
To investigate this counterintuitive result in more detail, small `structures' consisting of randomly placed Gaussian functions were used.  
For each combination of parameters, eight independent structure determination runs were performed, and for each run the achieved resolution was determined. 
The structure weights $w_i$ where chosen to be uniform and kept fixed during the simulated annealing runs.

\begin{figure}[t]
    \centering
    \includegraphics[width = \textwidth]{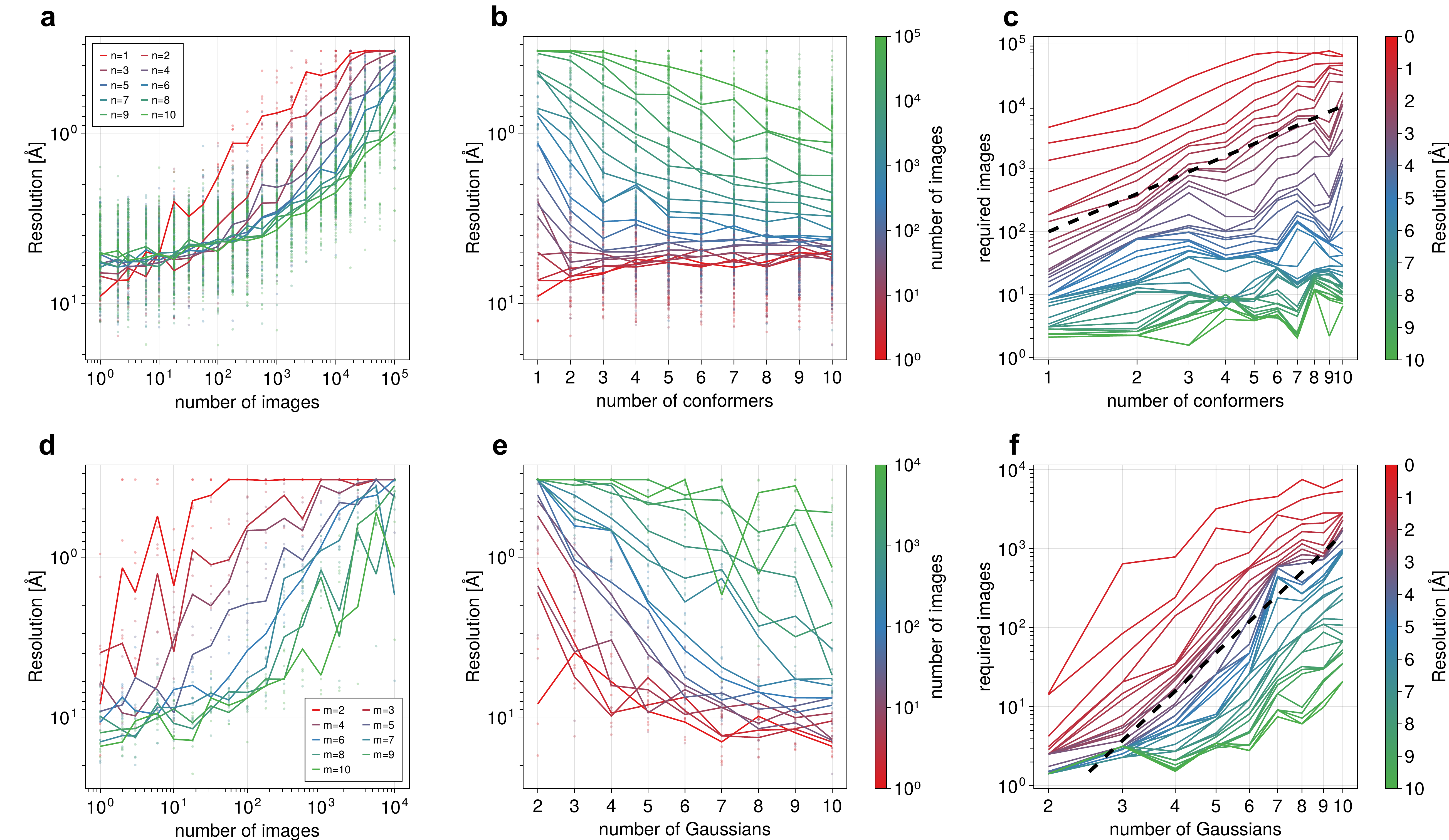}
    \caption{Dependence of the resolution on the number of images, the number of conformations, and the size of the structure. \textbf{a} Resolution as a function of the number of images for various numbers of conformations $n$. \textbf{b} Resolution as a function of the number of conformations for various numbers of images. \textbf{c} Required number of images to achieve various resolutions as a function of the number of conformations. For comparison, a quadratic relationship is shown (dashed line). \textbf{d} Resolution as a function of the number of images for various structure sizes (parameterized by the number of Gaussians $m$). \textbf{e} Resolution as a function of the number of Gaussians for various numbers of images. \textbf{f} Required number of images to achieve various resolutions as a function of the number of Gaussians. For comparison, a power law $m^5$ is shown (dashed line). }
    \label{fig: scaling}
\end{figure}

Figure~\ref{fig: scaling}c and~\ref{fig: scaling}f show for each combination of parameters the smallest number of images for which all of the replicas achieved a resolution better than a given threshold. 
As can be seen in Fig.~\ref{fig: scaling}c, for the structure ensemble of $n$ conformations with $m$ residues each, the required number of images is approximately proportional to $n^2$, the square of the number of conformations. 
This finding is in line with a theoretical argument showing that the information content of a single image is in this case proportional to $1/n^2$ (Supplementary Note 2). 
In contrast, the number of images required to resolve a single structure of $n\times m$ residues grows even much faster, approaching a power law $m^c$ with an exponent $c \approx 5$ for increasing resolution (Fig.~\ref{fig: scaling}f). Hence, for given complexity, ensemble refinement seems to be easier than single structure determination.

\clearpage
\section{Discussion}
Here we have developed a rigorous Bayesian method for determining biomolecular structures from single molecule X-ray scattering images in the extreme few photon Poisson regime. 
Using synthetic scattering images generated from simulated X-ray scattering experiments, we have demonstrated that both single structures as well as structural ensembles of small biomolecules can be resolved to near atomic resolution. 

Our results for the globular protein crambin show that a similar resolution of $4.2\,${\AA} is obtained compared to previous correlation based methods~\cite{von_ardenne_structure_2018} which also require very few photons per scattering image. Because such correlation based methods disregard higher correlations, whereas the full information content of each image is used in our Bayesian approach, the latter should require fewer images. 
This was indeed observed for the above protein, for which the number of images required to obtain near atomistic resolution was reduced from roughly $2\cdot 10^8$ to $1\cdot 10^8$. Assuming a pulse rate of 27,000 per second~\cite{zastrau_high_2021} and a $10\%$ hit fraction, this would reduce required beam time from 20 to 10 hours.

Because the rather small test proteins studied here scatter very few photons, they are conceptually more challenging than larger proteins~\cite{walczak_bayesian_2014}. For the latter, in contrast, the main bottleneck is computational cost, which will need to be addressed by improved optimization or sampling methods or by utilizing prior structural information, either from structure databases, from AlphaFold~\cite{jumper_highly_2021}, or guided by molecular dynamics force fields.

For alanine dipeptide the full conformational ensemble generated by an atomistic simulation was extracted at atomistic resolution of on average $1.8\,${\AA} from simulated scattering experiments, in which not only the current orientation of the biomolecule but also its current conformer was unknown.
For the 10 amino acid protein chignolin~\cite{honda_crystal_2008} both the folded and unfolded ensembles were resolved, albeit so far at lower resolution. 
Notably, also the weights corresponding to the folded and unfolded conformers were accurately recovered. 
Using weighted ensembles allows the number of conformers to be determined dynamically, as demonstrated by vanishing weights for incorrect structure poses.

Unexpectedly few images were required to resolve structural ensembles.
Because an ensemble of $n$ conformers consisting of $m$ atoms each has the same number of degrees of freedom as a single structure of $n\times m$ atoms, a similar number of images should be required. 
However, closer analysis suggests that roughly $O(m^5)$ images are required to a resolve a single structure with $m$ atoms. One might therefore expect that $O(n^5m^5)$ images are required for an ensemble of $n$ such structures. However, our test refinements suggest that only $O(n^2m^5)$ images are required, consistent with an expected information content of $O(1/n^2)$ for single scattering image.
This result suggests that in terms of the required number of scattering images, even determining more complex conformational ensembles should be possible with current experimental technology. 

Our Bayesian analysis of structural heterogeneity is similar in spirit to approaches that were successfully applied in cryo-electron microscopy~\cite{cossio_bayesian_2013,grant_cistem_2018,punjani_cryosparc_2017,lyumkis_likelihood-based_2013,scheres_chapter_2016,tang_eman2_2007,zivanov_new_2018,kinman_uncovering_2022}, which, from a mathematical standpoint, shares some similarities with single molecule X-ray scattering, albeit at a much lower noise level. From a more general perspective, our Bayesian approach represents a systematic and rigorous approach to include shot noise in the extreme Poisson regime characteristic for single molecule X-ray scattering experiments. 
In contrast to other proposed methods, this Bayesian framework will also allow to include other sources of noise and uncertainty in a conceptually straightforward manner, such as incoherently scattered photons, background scattering, detector noise, or scattering by disordered water at the biomolecular surface. Proper inclusion of these experimental uncertainties in terms of calibrated forward noise models (e.g., as documented in Ref.~\cite{yoon_comprehensive_2016}) will be the next crucial step towards atomistically resolved multiple structure ensembles. If successful, and in contrast to diffraction experiments on nano-crystals, this single molecule approach might ultimately provide a route to time resolved structures --- molecular movies --- without the need for synchronization through optical laser pulses.

\section{Methods}

\paragraph{Structure and structure ensemble representation.}
\label{sec: representation}
Electron density functions of the reference structures or conformers were described by a sum of $m$ of Gaussian functions with atomic positions $\vec y_i$, heights $h_i$ and standard deviations $\sigma_i$,
\begin{equation}\label{eq: representation}
    \rho(\vec r) = \sum_{i=1}^m \frac{h_i}{\left(\sigma_i\sqrt{2\pi}\right)^3} \exp \left(\frac1{2\sigma_i^2}\lVert \vec r - \vec y_i\rVert^2\right)\,.
\end{equation}
Electron density functions of the determined structures were described similarly, with one common height $h = h_i$ and one common standard deviation $\sigma=\sigma_i$, which is treated as an unknown and determined together with the positions $\vec y_i$.
Structural ensembles were represented by a weighted sum of conformers $\boldsymbol\uprho = \{\rho_1, \dots, \rho_n\}$ with weights $\vec w = \{w_1, \dots, w_n\}$.

\paragraph{Synthetic data generation.}
\label{sec: datagen}

For each of the synthetic scattering images, the photon positions on the detector $D$ were drawn from a probability distribution proportional to the intensity function $I(\vec k) = |F\{\rho\}(\vec k)|^2$ restricted to the appropriate Ewald sphere.
Specifically, generation of each image involved the following steps:
\begin{enumerate}
    \item A conformation of the molecule is selected randomly from the reference ensemble (for example, consisting of molecular dynamics trajectories),
    \item a random orientation $\vec R$ of the molecule is drawn uniformly from the rotation group $\mathrm{SO}(3)$,
    \item the number of scattered photons is drawn from a Poisson distribution with mean $N\int_D I(\vec R\vec k) \,\mathrm d \vec k$, where $N$ is the incoming beam intensity,
    \item the position of each scattered photon is drawn from the probability distribution proportional to $(I\circ \vec R)|_D$.
\end{enumerate}
The last two steps were implemented using rejection sampling. To this end, a von Mises-Fisher distribution $p$ on $D$ was chosen with high enough standard deviation that $I(\vec k) \leq p(\vec k)$ everywhere. Then, for each photon, its position $\vec k$ was drawn from $p$ and it was accepted with probability $I(R\vec k) / p(R\vec k)$. The beam intensity $N$ was chosen together with a normalization of $\rho$ such that this procedure accurately produces a Poisson distribution of the desired expected number of photons per image.

\paragraph{Computation of likelihoods.}
\label{sec: distribution}
The probability density of observing an image defined by photon positions $\vec k_1, \dots, \vec k_l$ given an electron density function $\rho$ with is corresponding intensity function $I(\vec k)$ was computed by averaging over all possible orientations $\vec R\in \mathrm{SO}(3)$ of the molecule, 
\begin{equation}
\label{eq: probability density}
    \begin{aligned}
        P(\vec k_1, \dots, \vec k_l \,|\, \rho)
        &= \frac{N^l}{l!}\int_{\mathrm{SO}(3)} \exp\left(-N\!\int_D I(\vec R\vec k) \,\mathrm d \vec k\right) \left(\prod_{i=1}^l I(\vec R\vec k_i)\right) \,\mathrm d \vec R, \\
    \end{aligned}
\end{equation}
where for each orientation, the probability is a product of the Poisson distribution for the number of photons $l$ in the image and a factor depending on the photon positions. 
These integrals were approximated by averaging over a discrete set of typically $r\approx 10^3$ to $r\approx 10^5$ rotations $\vec R_i$ with weights $s_i$,
\begin{equation}
    P(\vec k_1, \dots, \vec k_l \,|\, \rho) \approx \frac{N^l}{l!}\sum_{i=1}^r s_i\exp\left(-N\!\int_D I(\vec R_i\vec k) \,\mathrm d \vec k\right) \prod_{j=1}^l I(\vec R_i\vec k_j).
\end{equation}
The rotations $\vec R_i$ and their weights $s_i$ are constructed by combining a Lebedev quadrature rule on $S^2$ with a uniform quadrature rule on $S^1$ via the Hopf map~\cite{noauthor_sphere_lebedev_rule_nodate,graf_sampling_2009} (Supplementary Note~3).

\paragraph{Simulated annealing and hierarchical sampling.}
\label{sec: annealing}
A Monte Carlo simulated annealing approach with the energy function $-\log P$ was used to sample from or maximize the Bayesian posterior probability, as described in detail in the Supplement.
To enhance convergence, Bayesian sampling and maximization were performed in multiple hierarchical resolution stages. 
Starting from a low resolution representation of $\rho$ with correspondingly few degrees of freedom, the number of Gaussian functions was doubled in each stage and the reduced resolution structure determined by the previous stage was used as a proposal density (see Supplement). 
To calculate likelihoods for the reduced resolution structures, lower resolution scattering images were generated from the original images by rejection sampling, that is, by removing each photon in the original images with probability $1 - \exp(-\sigma^2k^2/2)$.
By construction, this rejection scheme samples from a Fourier transformed density $I_\rho \cdot \exp(-\sigma^2|\vec k|^2/2)$ which, by the convolution theorem, corresponds to a smoothed real space density $\tilde\rho = \rho*\mathcal N(\sigma)$ obtained as the convolution of $\rho$ with a Gaussian kernel with width (resolution) $\sigma$. 
Computational efficiency was further increased substantially by selecting only those original images for the likelihood computations that actually contain useful information at the respective resolution. As described in the Supplement, the Bayesian formalism allows for removing this selection bias.

\paragraph{Structure alignment and resolution estimate.}
Because the orientations of the obtained structures are irrelevant, these were rotationally aligned to each other
by minimizing the cost function
\begin{equation}
    d(\vec S) = \frac1n \sum_{i=1}^n \min_{j=1}^m\, \lVert\vec y_i - \vec S \vec y_j'\rVert + \frac1m \sum_{j=1}^m \min_{i=1}^n\, \lVert\vec y_i - \vec S \vec y_j'\rVert.
\end{equation}
Here, the positions $\vec y_1, \dots, \vec y_n$ and $\vec y_1', \dots, \vec y_m'$ define two structures per equation \eqref{eq: representation} and $\vec S$ is a rotation matrix $\vec S \in \mathrm{O}(3)$. Both rotations and reflections were included, as X-ray scattering images do not distinguish between mirror images. 

The resolution of the aligned structures was estimated using Fourier shell correlations~\cite{van_heel_fourier_2005},
\begin{equation}
    \mathrm{FSC}(k) = \frac{\int_{\lVert\vec k\rVert = k} \hat\rho_1(\vec k)^*\hat\rho_2(\vec k)\,\mathrm d\vec k}{\sqrt{\int_{\lVert\vec k\rVert = k} |\hat\rho_1(\vec k)|^2\,\mathrm d\vec k}\sqrt{\int_{\lVert\vec k\rVert = k} |\hat\rho_2(\vec k)|^2\,\mathrm d\vec k}}\,,
\end{equation}
where $\rho_1$ are $\rho_2$ the structures to be compared and $\hat\rho$ denotes the Fourier transform of $\rho$. Accordingly, the achieved resolution was determined as $2\pi/k_\mathrm{fsc}$, where $k_\mathrm{fsc}$ is the threshold at which the Fourier shell correlation drops below $1/2$, providing a conservative estimate~\cite{van_heel_fourier_2005}.

\paragraph{Molecular dynamics simulations.}
All atomistic simulation trajectories were generated using the GROMACS 2018 software package~\cite{abraham_gromacs_2015} with the Charmm36mm force field~\cite{huang_charmm36m_2017} and the OPC water model~\cite{izadi_building_2014}. 
For chignolin, the starting structure was taken from the Protein Data Bank~\cite{berman_protein_2000}, entry 5AWL~\cite{honda_crystal_2008}.
All hydrogen atoms were described by virtual sites~\cite{feenstra_improving_1999}.
Each protein was placed within a triclinic water box, such that the smallest distance between protein surface and box boundary was larger than $1.5\,$nm. 
Sodium and chloride ions were added to neutralize the system, corresponding to a physiological concentration of $150\,$mmol/l. 
Energy minimization was performed using steepest descent for $5\cdot 10^4$ steps.
Each system was subsequently equilibrated for $0.5\,$ns in the $NVT$ ensemble, and subsequently for $1.0\,$ns in the $NPT$ ensemble at $1\,$atm pressure and temperature $300\,$K using an integration time step of $2\,$fs. 
The velocity rescaling thermostat~\cite{bussi_canonical_2007} and Parrinello-Rahman pressure coupling~\cite{parrinello_polymorphic_1981} were used with coupling coefficients of $\tau=0.1\,$ps and $\tau=1\,$ps, respectively.
All bond lengths of the solute were constrained using the LINCS algorithm~\cite{hess_lincs_1997} with an expansion order of 6, and the geometry of the water molecules was constrained using the SETTLE algorithm~\cite{miyamoto_settle_1992}. 
Electrostatic interactions were calculated using PME~\cite{darden_particle_1993}, with a real space cutoff of $10\,${\AA} and a Fourier spacing of $1.2\,${\AA}.
For all production runs, a $4\,$fs integration was used, and the atom coordinates were saved every $100\,$ps, such that $10^5$ snapshots were available for each trajectory.
The trajectories for alanine dipeptide were taken from mdshare \cite{noauthor_mdshare_nodate-1}. The structure for crambin was taken from PDB entry 1EJG~\cite{jelsch_accurate_2000}.

\paragraph*{Data Availability.}
All relevant data are available from the authors.

\paragraph*{Code Availability.}
We have implemented our method in the Julia programming language \cite{bezanson_julia_2017}. The source code is available at \href{https://gitlab.gwdg.de/sschult/xfel}{\url{https://gitlab.gwdg.de/sschult/xfel}}.

\printbibliography

\section*{Acknowledgments}
This work was financially supported by the Federal Ministry of Education and Research through the joint research project 05K20EGA Fluctuation XFEL, and the Deutsche Forschungsgemeinschaft (DFG, German Research Foundation) - CRC 1456/1 - 432680300. The MD-trajectories were kindly provided by Nicolai Kozlowski. 

\section*{Author contributions}
S.S. and H.G. conceived research, S.S. carried out research, S.S. and H.G. wrote paper. 

\section*{Competing interests}
The authors declare no competing interests.

\begin{appendix}
\appendixpage
\renewcommand{\thesection}{\arabic{section}}

\section{Parameters}
\label{sup: parameters}

The parameters used for the test cases are shown in Table~\ref{tab: parameters}.
The Lebedev precision and the number of angular rotations $S_j$ are chosen such that the expected angular distance between nearest neighbors in the resulting grid is smaller than the length scale corresponding to the desired relative resolution divided by the approximate radius of the molecule. 
Due to hardware constraints, the number of angular rotations must be a multiple of $32$. The parameters for image selection ($r_i$ and $m_i$) were chosen such that the radial distribution of photons in the selected images was close to uniform up to the desired resolution level.

\addtolength{\tabcolsep}{-2.1pt}

\begin{table}[ht!]
    \footnotesize
    \centering
    \pgfplotstabletypeset[
        col sep=&, row sep=\\,
        every head row/.style={before row = \toprule, after row = \midrule},
        every last row/.style={after row = \bottomrule},
        every row no 4/.style={after row = \midrule},
        every row no 6/.style={after row = \midrule},
        every row 6 column 3/.style={string type},
    	every row 7 column 3/.style={string type},
    	columns/name/.style={column name = {Name}, string type},
    	columns/stage/.style={column name = {Stage}},
    	columns/total/.style={column name = {\shortstack{total\\images}}, sci, sci zerofill, precision=2},
    	columns/selected/.style={column name = {\shortstack{selected\\images}}, fixed},
    	columns/nmask/.style={column name = {$n_i$}, string type},
    	columns/rmask1/.style={column name = {$r_i\,\,[1/\text{{\AA}}$]}, string type},
    	columns/rmask2/.style={column name = {$b_i$}, string type},
    	columns/kwidth/.style={column name = {$\sigma\,\,[\text{{\AA}}]$}, fixed, fixed zerofill, precision = 1},
    	columns/T0/.style={column name = {$T_0$}},
    	columns/thalf/.style={column name = {$t_{1/2}$}, sci, precision=1},
    	columns/ngauss/.style={column name = {$m$}},
    	columns/nstrucs/.style={column name = {$n$}},
    	columns/precision/.style={column name = {{\shortstack{Lebedev\\precision}}}},
    	columns/npercircle/.style={column name = {{\shortstack{angular\\rotations}}}},
    ]{
    	name & stage & total & selected & nmask & rmask1 & kwidth & thalf & ngauss & nstrucs & precision & npercircle \\
    	\multirow{5}{*}{Crambin} & 1 & 8956 & 1000 & (4) & $(0.25, \infty)$ & 2.0 & 1000 & 12 & 1 & 23 & 32 \\
    	& 2 & 1e7 & 19315 & (3, 2) & $(0.33, 0.5, \infty)$ & 1.5 & 10000 & 23 & 1 & 47 & 32 \\
    	& 3 & 3044813 & 50000 & (1, 1, 2) & $(0.35, 0.5, 0.65, \infty)$ & 1.2 & 20000 & 46 & 1 & 47 & 64 \\
    	& 4 & 1e8 & 204447 & (1, 2, 2) & $(0.35, 0.5, 0.8, \infty)$ & 0.9 & 100000 & 92 & 1 & 89 & 64 \\
    	& 5 & 1e8 & 634032 & (1, 1, 3) & $(0.4, 0.65, 0.9, \infty)$ & 0.5 & 100000 & 184 & 1 & 89 & 64 \\
        \multirow{2}{*}{Dipeptide} & 1 & 1000000 & 3965 & (4, 4) & $(0.9, 1.3, \infty)$ & 0.5 & 1e3 & 10 & 2 & 23 & 32 \\
        & 2 & 1000000 & - & - & - & 0.0 & 5e3 & 10 & 8 & 35 & 32 \\
    	\multirow{3}{*}{Chignolin} & 1 & 10000 & - & - & - & 2.5 & 1000 & 5 & 2 & 23 & 32 \\
    	& 2 & 10931255 & 100000 & (2, 2) & $(0.4, 0.6, \infty)$ & 1.5 & 5000 & 10 & 4 & 23 & 32 \\
    	& 3 & 12382653 & 100000 & (2, 3) & $(0.4, 0.6, \infty)$ & 1.2 & 10000 & 20 & 6 & 47 & 64\\
    }
    \caption{Parameters for the three test cases.}
    \label{tab: parameters}
\end{table}

\section{Expected information content of scattering images}
\label{sup: information}
The information content of scattering image on structural ensembles can be estimated analogous to an argument for mixtures of normal distributions \cite{behboodian_information_1972}, as follows. 
Consider an ensemble of two structures $\rho_1$ and $\rho_2$ with weights $w$ and $1-w$, respectively. The probability of observing an image $x$ is then a mixture of the two single distributions,
\begin{equation}
    p(x; \rho_1, \rho_2) = w p(x; \rho_1) + (1 - w) p(x; \rho_2).
\end{equation}
By the Bayesian central limit theorem, in the limit of many scattering images the posterior becomes a multivariate normal distribution with covariance $N^{-1}I^{-1}$, 
\begin{equation}
    P(\rho_1, \rho_2\,|\, \mathcal I) \approx \mathcal N(\rho_1, \rho_2;  N^{-1}I^{-1}),
\end{equation}
where $N$ is the number of images, and $I$ the Fisher information matrix. 
The first diagonal element of this matrix is approximately proportional to the weight squared,
\begin{equation}
    I_{\rho_1\rho_1} = \operatorname{E}\!\left[\left(\frac{\partial}{\partial{\rho_1}} \log p(x; \rho_1, \rho_2)\right)^{\!\!2}\right]
        = \operatorname{E}\!\left[\left(\frac{w \frac{\partial}{\partial\rho_1} p(x; \rho_1)}{p(x; \rho_1, \rho_2)}\right)^{\!\!2}\right]
        = w^2 \,\operatorname{E}\!\left[\left(\frac{\frac{\partial}{\partial\rho_1} p(x; \rho_1)}{p(x; \rho_1, \rho_2)}\right)^{\!\!2}\right].
\end{equation}
Therefore, under the assumption that the off-diagonal elements are small, the limiting variance for $\rho_1$ becomes $1/(N w^2)$. 
An similar argument can be carried out for more than two distinct structures. 
In the special case of uniform weights $w=1/n$ the limiting variance becomes $n^2/N$, consistent with the quadratic scaling observed in Fig.~5.

\section{Computation}
\label{sup: computation}

The integral over $\mathrm{SO}(3)$ is approximated by a finite sum over rotations $\vec R_i$ with weights $s_i$,
\begin{equation}
    P(\vec k_1, \dots, \vec k_n \,|\, \rho) \approx \frac{N^n}{n!}\sum_i s_i\exp\left(-N\int_D I(\vec R_i\vec k) \,\mathrm d \vec k\right) \prod_{j=1}^n I(\vec R_i\vec k_j)
\end{equation}
Computing this sum involves evaluating the intensity function $I$ at all points of the form $\vec R_i\vec k_j$.
Since this has to be done for all the images, this leads to a very large number of evaluations of $I$. 
It is therefore efficient to first discretize the images. 
To that end, the detector is pixelated, that is, partitioned into a grid of cells with centers $\vec x_k$ and areas $a_k$. Each image $\vec k_1, \dots, \vec k_n$ is replaced with a set of indices $k_1, \dots, k_n$, such that for each $\vec k_i$ the closest point in the grid is $\vec x_{k_i}$. In this setting, the probability distribution becomes
\begin{equation}
    \label{eq:grids}
    P(k_1, \dots, k_n \,|\, \rho) \approx \frac{N^n}{n!}\sum_i s_i \exp\left(-N\sum_k a_k I(\vec R_i\vec x_k)\right) \prod_{j=1}^n a_{k_j} I(\vec R_i\vec x_{k_j})
\end{equation}

To construct the quadrature rule for $\mathrm{SO}(3)$, we proceed as follows. First, we choose a Lebedev grid as a uniform grid of points $\vec v_i$ in the 2-sphere $S^2$. For each one of these, we find a rotation $Q_i \in \mathrm{SO}(3)$ such that $Q_i\vec v_i \parallel \vec k_0$. In addition, let $S_j$ be uniformly spaced rotations around the axis defined by $\vec k_0$. The set of products $S_jQ_i$ is then a uniform grid in $\mathrm{SO}(3)$. Equation \eqref{eq:grids} becomes
\begin{equation}
    P(k_1, \dots, k_n \,|\, \rho) \approx \frac{N^n}{n!}\sum_{i,j} s_i\exp\left(-N \sum_k a_k I(Q_i S_j\vec x_k)\right) \prod_{m=1}^n a_{k_m} I(Q_iS_j\vec x_{k_m})
\end{equation}
Choosing the pixel grid $\vec x_k$ such that it is rotationally symmetric allows further simplification. We reindex it as $\vec x_{k,l}$, such that $S_j\vec x_{k,l} = \vec x_{k+j,l}$. Here, the first index is considered cyclic, that is, if, say, $k$ ranges from $1$ to $k_\mathrm{max}$, then $\vec x_{k+j,l}$ is to be interpreted as $\vec x_{(k+j\mod k_\mathrm{max}),l}$. The corresponding areas $a_{k,l}$ only depend on $l$, so we write $a_l = a_{k,l}$. The images now also consist of these new indices. Plugging this in, we get
\begin{align}
    P(k_1, l_1, \dots, k_n, l_n \,|\, \rho) &\approx \frac{N^n}{n!}\sum_{i,j} w_i\exp\left( -N \sum_{k,l} a_l I(Q_iS_j\vec x_{k,l})\right) \prod_{m=1}^n a_{l_m} I(Q_iS_j\vec x_{k_m,l_m})\\
    &= \frac{N^n}{n!}\sum_i w_i\exp\left( -N \sum_{k,l} a_l I(Q_i\vec x_{k,l})\right) \sum_j\prod_{m=1}^n a_{l_m} I(Q_i\vec x_{k_m+j,l_m}) \\
    &= \frac{N^n}{n!}\sum_i w_i P_i \sum_j\prod_{m=1}^n I_{i,k_m+j,l_m}
\end{align}
The values $I_{i,k,l} \coloneqq a_lI(Q_i\vec x_{k,l})$ and $P_i \coloneqq \exp(-N\sum_{k,l} I_{i,k,l})$ can be computed in advance and reused for each image.

Due to limited floating point precision, a number of adjustments must be made. Due to the large value of $N$, computing $P_i$ results in underflow. Therefore, we write \begin{equation}
    \tilde P_i = P_i /\bar P, \qquad \bar P =\left(\prod_{i'=1}^{i_\mathrm{max}} P_{i'}\right)^{\frac1{i_\mathrm{max}}}.
\end{equation}
Further, $I_{i,k,l} \ll 1$, so if the images contain enough photons the product over $m$ will underflow. Since the magnitude of $I_{i,k,l}$ depends mostly on $l$, we define
\begin{equation}
    \tilde I_{i,k,l} = I_{i,k,l} /\bar I_l, \qquad \bar I_l = \frac1{i_\mathrm{max}k_\mathrm{max}}\sum_{i'=1}^{i_\mathrm{max}}\sum_{k'=1}^{k_\mathrm{max}} I_{i',k',l}
\end{equation}
Both $\bar P$ and $\bar I_l$ do not depend on the rotation index $i$ and factor out,
\begin{equation}
    P(k_1, l_1, \dots, k_n, l_n \,|\, \rho) \approx \frac{N^n}{n!}\bar P\left(\prod_{m=1}^n\bar I_{l_m}\right)\sum_i w_i \tilde P_i \sum_j\prod_{m=1}^n \tilde I_{i,k_m+j,l_m}
\end{equation}
Taking the logarithm,
\begin{equation}
    \log P(k_1, l_1, \dots, k_n, l_n \,|\, \rho) \approx \log \frac{N^n}{n!} + \log\bar P + \sum_{m=1}^n\log \bar I_{l_m}
    +\log \sum_i w_i \tilde P_i \sum_j\prod_{m=1}^n \tilde I_{i,k_m+j,l_m},
\end{equation}
we see that only $\log \bar P$ and $\log \bar I_l$ appear, which can be computed without overflow.

\section{Monte Carlo Simulated Annealing}
\label{sup: annealing}

Let $\boldsymbol\uprho = (\rho_1, \dots, \rho_n)$ and $\vec w = (w_1, \dots, w_n)$ denote vectors of electron densities and weights, respectively. 
A Markov chain of structural ensembles $\boldsymbol\uprho_t$ with weights $\vec w_t$ was constructed iteratively using a Metropolis-within-Gibbs algorithm. This algorithm works as follows.
For each step $t$, first a Metropolis step for the structures is performed, that is, new candidate structures $\boldsymbol\uprho'$ are drawn from a proposal distribution $g(\boldsymbol\uprho' | \boldsymbol\uprho_t)$, and this candidate is accepted ($\boldsymbol\uprho_{t+1} = \boldsymbol\uprho'$) or rejected ($\boldsymbol\uprho_{t+1} = \boldsymbol\uprho_t$) with probability 
\begin{equation}
    1 \wedge \exp\left(\frac{\log P(\boldsymbol\uprho', \vec w_t \,|\, \mathcal I) - \log P(\boldsymbol\uprho_t, \vec w_t \,|\, \mathcal I) + \log g(\boldsymbol\uprho_t | \boldsymbol\uprho') - \log g(\boldsymbol\uprho' | \boldsymbol\uprho_t)}{T(t)}\right),
    \label{eq: metropolis}
\end{equation}
adopting the notation $1 \wedge x = \min(1, x)$.
The temperature $T(t)$ is determined according to an exponential annealing schedule $T(t) = T_0\exp(-\lambda t)$ for some constant $\lambda$. 
The proposal density $g$ is an isotropic normal distribution $\mathcal N(\boldsymbol\uprho_t, d)$ around $\boldsymbol\uprho_t$, that is, to obtain the candidate, the position of each Gaussian in the structure representation is perturbed by a normally distributed amount; or it is given by our hierarchical sampling method as described in the next section.
The step size $d$ is determined iteratively such that the acceptance rate is the optimal $23\%$~\cite{gelman_weak_1997}, by increasing or decreasing it after a successful or unsuccessful step, respectively.

Second, a separate Metropolis step for the weights is performed. To correctly sample from the $n$-simplex of weights $w_i$ such that $w_i \leq 0$ and $\sum_i w_i = 1$, we introduce variables $s_j \leq 0$ such that $w_i = s_i / \sum_j s_j$. 
For these variables, the proposals are drawn from a Gamma distribution of mean $s_j$ and standard deviation given by the current step size. 
Note that this is not a proposal distribution in the sense of equation \eqref{eq: metropolis}, as it does not appear in the acceptance probability.
If one of the weights $w_i$ becomes zero during the sampling process, the corresponding structure $\rho_i$ does no longer affect the posterior probability, hindering convergence. 
To prevent this, a delayed acceptance scheme is used as follows.
Each proposal $\vec s'$ with corresponding weights $\vec w'$ generated by the above procedure is accepted with probability 
\begin{equation}
    g^*(\vec w'\,|\,\vec w_t) = 1 \wedge \exp\left(\frac{1}{2\nu}\lVert\vec w' - \vec c\rVert^2 - \frac{1}{2\nu}\lVert\vec w_t - \vec c\rVert^2\right),
\end{equation}
where $\vec c = (1/n, \dots, 1/n)$ and $\nu$ is sufficiently small to ensure that the weights remain non-zero.
Finally, the proposal is accepted with probability 
\begin{equation}
    1 \wedge \exp\left(\frac{\log P(\boldsymbol\uprho_{t+1}, \vec w' \,|\, \mathcal I) - \log P(\boldsymbol\uprho_{t+1}, \vec w_t \,|\, \mathcal I) + \log g^*(\vec w_t | \vec w') - \log g^*(\vec w' | \vec w_t)}{T(t)}\right).
\end{equation}
The metropolis step for the weights has little computational cost, as the computationally costly parts of equation \eqref{sup: computation} are unaffected. 
Therefore, it is repeated multiple times in each iteration.

\section{Proposal density for hierarchical sampling}
\label{sup: proposal}
In each hierarchical sampling stage, the number of Gaussian functions was doubled, and the reduced resolution structure determined by the previous stage was used as a proposal density to improve convergence in the simulated annealing, as follows. 
Let $\vec y_1, \dots, \vec y_n$ be the positions of the Gaussian functions from the previous stage, and $\vec z_1, \dots, \vec z_{2n}$ those of the current stage. 
Then the proposal density was, up to normalization, given by
\begin{equation}\label{eq: proposal density}
    g(\vec z_1', \dots, \vec z_{2n}' \,|\, \vec z_1, \dots, \vec z_{2n}) \propto \prod_{i=1}^{2n} \exp\!\left(-\frac{\lVert\vec z_i' - \vec z_i\rVert^2}{2\sigma^2}\right) \prod_{i=1}^{n} \exp\!\left(-\frac{\lVert\vec z_{2i}' - \vec y_i\rVert^2 + \lVert\vec y_{2i+1}' - \vec y_i\rVert^2}{2w^2}\right),
\end{equation}
where $w$ is the width of the Gaussians from the previous stage.
For ensembles of structures, the proposal density becomes a product over the single structures $\rho_i$ with separate intermediates for each $\rho_i$, 
\begin{equation}
    g(\boldsymbol\uprho'\,|\,\boldsymbol\uprho_t) = \prod_{i=1}^n g(\rho_i'\,|\,\rho_i),
\end{equation}
where $g(\rho_i'\,|\,\rho_i)$ is the proposal density from equation \eqref{eq: proposal density}.

\section{Image selection}
\label{sup: selection}
In our hierarchical sampling scheme, images containing only photons with $|\vec k|$ below a threshold are no longer useful, and the computations were sped up by removing these images. 
To achieve this, numbers $(r_i)$ and integers $m_i$ were chosen, and only the subset $\mathcal I_C$ of those images was used that fulfilled the condition $C(I)$ that for each $i$ the image $I$ contains at least $m_i$ photons with $r_i < |\vec k| < r_{i+1}$. 
To ensure that the posterior was not biased by this filtering, it was taken into account in the Bayesian formalism by dividing by the probability $P(C\,|\,\boldsymbol\uprho, \vec w)$ that an image fulfills $C$. In other words, the original posterior probability was replaced with $P(\boldsymbol\uprho, \vec w\,|\,\mathcal I_C, C) \propto P(\mathcal I_C \,|\, \boldsymbol\uprho, \vec w) / P(C\,|\,\boldsymbol\uprho, \vec w)$. The probability that an image fulfills $C$ depends on both the orientation $\vec R$ and the conformer $i$. Therefore, $P(C\,|\,\boldsymbol\uprho, \vec w)$ was obtained by averaging over both,
\begin{equation}
    P(C\,|\,\boldsymbol\uprho, \vec w) = \sum_j w_j \int_{\mathrm{SO}(3)} \prod_i \left(1 - Q\left(m_i-1, N\!\!\int_{D_i}\!\!\mathcal \lvert F\{\rho_j\}(\vec R\vec k)\rvert^2 \,\mathrm d \vec k \right)\right) \mathrm d \vec R,
\end{equation}
where $Q(x, \lambda)$ is the cumulative distribution function of a Poisson distribution with mean $\lambda$ and $D_i = \{\vec k \in D \,|\, r_i < \lVert \vec k\rVert < r_{i+1}\}$ is the relevant slice of the Ewald sphere.

\end{appendix}

\end{document}